\newcommand{\orcidX}[1]{\href{https://orcid.org/#1}{\includegraphics[width=10pt]{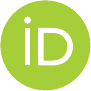}}}
\documentclass[sigconf,nonacm]{acmart} %double column

\AtBeginDocument{%
  \providecommand\BibTeX{{%
    \normalfont B\kern-0.5em{\scshape i\kern-0.25em b}\kern-0.8em\TeX}}}

\setcopyright{none}
\acmDOI{}

\usepackage{lscape}
\usepackage{multirow}
\usepackage[nolist,nohyperlinks]{acronym}
\usepackage{colortbl}
\usepackage{enumitem}

\begin{document}

\begin{acronym}

\acro{ML}{\emph{Machine Learning}}
\acro{SVM}{\emph{Support Vector Machine}}
\acro{KNN}{\emph{K-Nearest Neighbors}}
\acro{NB}{\emph{Naive Bayes}}
\acro{RF}{\emph{Random Forest}}
\acro{HHMM}{hierarchical hidden Markov model}
\acro{LR}{Logistic Regression}
\acro{XGB}{XGBoost}

\acro{LSP}{Logic Scoring of Preference}

\acro{DoS}{\emph{Denial of Service}}
\acro{DDoS}{\emph{Distributed Denial of Service}}
\acro{ELOOCV}{Experiment-wise Leave-one-out Cross-Validation}

\acro{TN}{\emph{True Negative}}
\acro{FN}{\emph{False Negative}}
\acro{TP}{\emph{True Positive}}
\acro{FP}{\emph{False Positive}}
\acro{CM}{Confusion Matrix}

\acro{TCP}{Transmission Control Protocol}
\acro{UDP}{User Datagram Protocol}

\acro{VMs}{\emph{Virtual Machines}}

\acro{LSP}{Logic Scoring of Preference}
      
\end{acronym}


\title{ONDA: ONline Database Architect}


\author{Nuno Laranjeiro \orcidX{0000-0003-0011-9901}}
\orcid{0000-0003-0011-9901}
\affiliation{%
  \institution{University of Coimbra, CISUC, DEI}
  %\city{Coimbra}
  \country{Portugal}
}
\email{cnl@dei.uc.pt}

\author{Alexandre Miguel Pinto \orcidX{0000-0003-0577-0939}}
\orcid{0000-0003-0577-0939}
\affiliation{%
  \institution{Signal AI}
  \city{London}
  \country{United Kingdom}
}
\email{alexandre.pinto@signal-ai.com}

\renewcommand{\shortauthors}{N. Laranjeiro and A. M. Pinto}

%\begin{abstract}

%\end{abstract}
\begin{abstract}

Database modeling is a key activity towards the fulfillment of storage requirements. Despite the availability of several database modeling tools for developers, these often come with associated costs, setup complexities, usability challenges, or dependency on specific operating systems. In this paper we present ONDA, a web-based tool developed at the University of Coimbra, that allows the creation of Entity-Relationship diagrams, visualization of physical models, and generation of SQL code for various database engines. ONDA is freely available at \textit{\url{https://onda.dei.uc.pt}} and was created with the intention of supporting teaching activities at university-level database courses. At the time of writing, the tool being used by more than three hundred university students every academic year.
\end{abstract}






\maketitle

\section{Introduction}

Database Management Systems have been evolving since the 60's \cite{hughes_what_1999}. At the time, only specific systems required some form of data storage, nowadays most applications need some kind of data storage and management mechanisms. Depending on the type of application, simple storage mechanisms like the use of files may be sufficient, others may require an in-memory database, a dedicated database server with persistent storage, or, for instance, a cloud storage service \cite{garciaMolina_database_2008, li_modernization_2023}.

\begin{figure*}[h!]
    \centering    \includegraphics[width=0.8\linewidth]{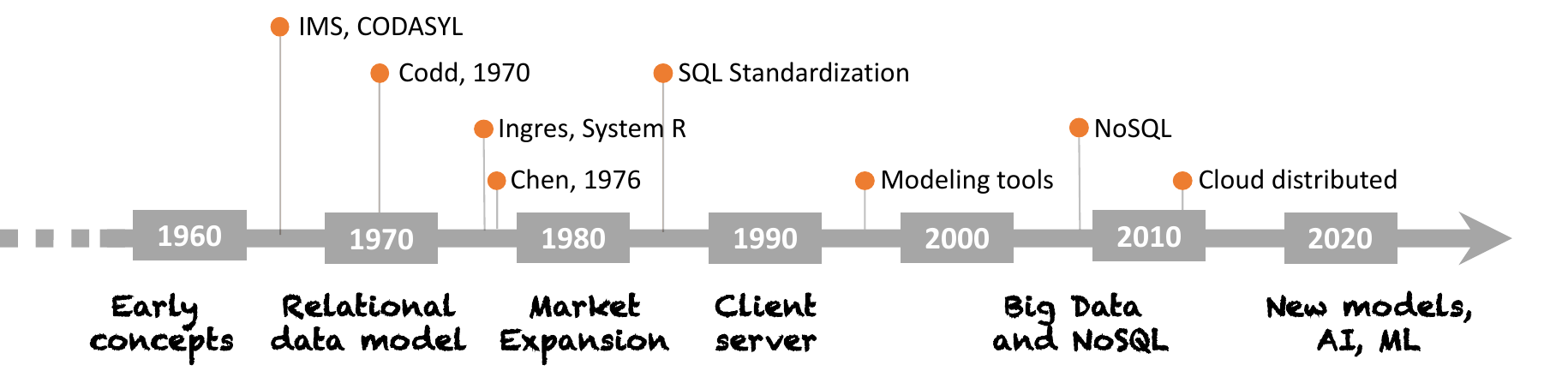}
    \caption{Timeline of advances in data storage and management.}
\label{timeline}
\end{figure*}

Data may be organized according to different models, which determine how data should be structured, and may be supported by different architectures, which determine how data is physically stored \cite{garciaMolina_database_2008}. Over time, the field of data storage and management has undergone numerous advancements \cite{hughes_what_1999, anthes_happy_2010}. Figure \ref{timeline} highlights some of the most relevant developments, in the context of this paper, which we describe in the following paragraphs.

In the 60's two data models were popular, namely a tree-structure model named IMS (1966) and a network model named CODASYL (1969) \cite{hughes_what_1999}. In the 70's \citeauthor{Codd1970} \cite{Codd1970} published a paper which proposes a relational model of data, separating the physical organization from the logical organization, i.e., the schema. A key idea behind the proposal was to safeguard users from disruptive changes in data representation as a consequence of data growth. The first relational database prototypes appeared in the mid 70's named Ingres \cite{stonebraker_design_1976} and System R \cite{astrahan_system_1976}.

 A new database model named Entity-Relationship (ER) was proposed by \citeauthor{chen1976} in \cite{chen1976}. The ER model is based on the definition of Entities and Relationships between them, which may be of various types (e.g., different cardinalities). The model includes some semantic information and can be viewed as a generalization or extension of the existing models at the time, or even as a framework from which other models (e.g., relational, entity set, or network models) could be derived \cite{chen1976}. In the 80's, in particular in 1986 the Structured Query Language (SQL) became an ANSI standard and also an ISO standard a few months later, acknowledging the importance of relational databases and also further fostering their popularity \cite{chamberlin_early_2012}.

In the 90's we witnessed significant advances in database systems and applications and in software for building database applications. Such software includes programming environments but also tools mostly centered on database modeling, such as Sybase Powerdesigner, whose origins trace back to 1995, ERwin Data Modeler in 1998, or Enterprise Architect a bit later \cite{dataModeling2011}.

A relational database named NoSQL was made available in 1998 and did not expose an SQL interface, which was usual at the time \cite{strauch2011nosql}. In the 00's, in light of the emergence of non-relational and distributed data stores (e.g., Google Bigtable, Amazon DynamoDB) Johan Oskarsson reintroduced the term NoSQL in 2009, as a way to label such storage systems \cite{strauch2011nosql}. The popularity of database systems that are using new architectures (e.g., document, key-value, graph, time series) has been growing in recent years, due to their overall effectiveness in particular contexts (e.g., the need for storing unstructured data). Despite such advances, the maturity and wide range of applications of relational databases make them a quite popular solution, among the various options available nowadays \cite{kaufmann_sql_2023}.

Tools for database modeling generally come associated with various limitations \cite{dataModeling2011}, as follows. Many times such tools are in the form of commercial software that needs to be paid in order to be used. Often the software is to be run locally and supports one or a reduced set of operating systems. The installation procedure may not be trivial and, depending on the license, the application may not allow for updates or upgrades. Especially important is the type of diagrams supported. Sometimes just the physical design of tables is supported, in some other times the tools provide a slighlty higher level of abstraction and it is possible to draw a logical representation of the data, but many tools do not allow (or have additional constraints, such as paid licenses) representing Entity-Relationship diagrams in the terms proposed by \cite{chen1976}.

\begin{figure*}[!h]
    \centering \includegraphics[width=1\linewidth]{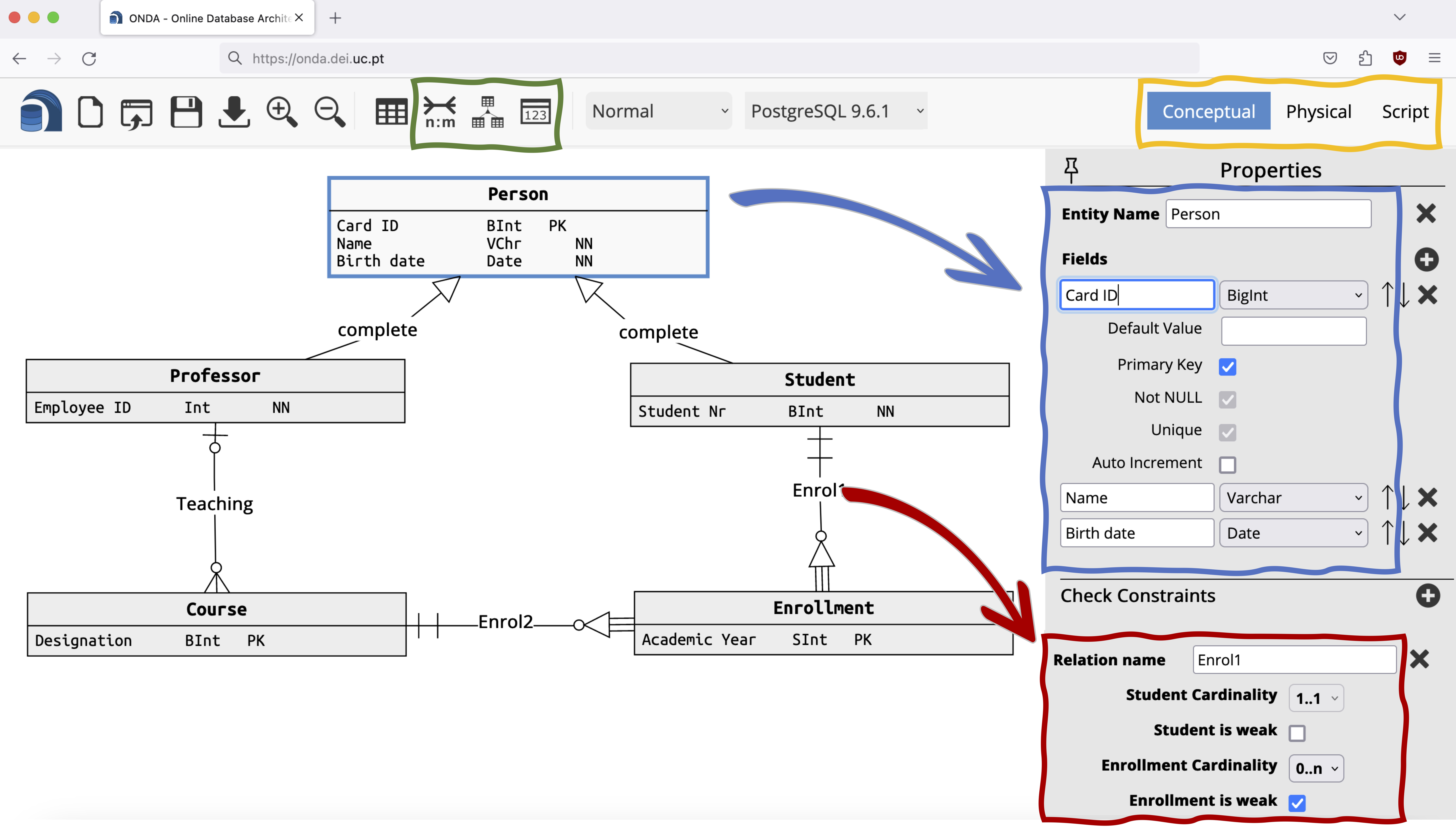}
    \caption{Main ONDA screen and the creation of an Entity-Relationship diagram.}
\end{figure*}

\textbf{In this paper we present the ONline Database Architect (ONDA), an open-source, web-based, and free tool for database modeling}. The concept of the tool has been created by the authors of this paper and the tool has been developed by several students of the Department of Informatics Engineering (DEI) of the Faculty of Sciences and Technology of the University of Coimbra in Portugal. The main idea behind the creation of the tool was to support various courses at DEI in which the creation of Entity-Relationship diagrams is required, while allowing students to have free access to a cross-platform tool that requires no setup. The tool runs via a browser and is available at \url{https://onda.dei.uc.pt}. As a quick overview, ONDA allows the following main functionalities:

\begin{itemize}
    \item Creation of Entity-Relationship diagrams;
    \item Conversion of Entity-Relationship diagrams to physical models;
\item Generation of SQL based on the generated physical models and supporting various database engines. 
\end{itemize}

ONDA has been the subject of various iterations and the current version (v4) is now being used by more than three hundred students every academic year and across various BSc and MSc courses. Despite a few existing limitations, it is sufficiently stable to be used throughout each academic semester and has received generally positive feedback from students. 

The paper is organized as follows. Section II presents the main functionalities provided by ONDA and Section III concludes the paper and identifies planned future work.
\section{The ONDA Application}

In this section, we overview the ONDA application, going through the following three main features: i) ER diagram creation; ii) physical model generation; and iii) SQL generation.

\subsection{Entity-Relationship Diagram Creation}

Figure 2 shows ONDA's main screen. In the top right-hand corner, marked with a yellow rectangle, we are able to select the following three options: i) 'Conceptual' - displays the screen shown in Figure 2, allowing the user to draw an Entity-Relationship diagram; ii) 'Physical' - generates and shows the physical diagram that corresponds to the ER created in the 'Conceptual' tab; and iii) 'Script' - generates the SQL corresponding to the created physical diagram.

The main drawing icons are marked with a green rectangle in Figure 2. From the left to the right, the first icon allows the placement of an entity on the canvas, the second icon allows to connect two entities with a relationship, and the last icon allows to define an inheritance relationship between two entities.

After \textbf{adding an Entity} to the canvas, it is possible to configure it by setting its name, adding attributes and configuring these attributes. This configuration is visible after selecting the Entity to configure and it is marked with a blue rectangle at the right-hand side of Figure 2. For each attribute it is possible to set the following properties: attribute name; datatype; datatype configuration (e.g., maximum number of characters if it is a string, precision and scale if it is a Number); identify it is a primary key; if it is mandatory; if it is unique; or if its value should be automatically incremented by the database management system. It is also possible to manually add SQL constraints (e.g., to specify the valid domain of a certain attribute), which are however not shown in Figure 2.

Marked with a red rectangle at the bottom right-hand side of Figure 2, is an overlay image to represent the case where the user wants to \textbf{configure a relationship}. Notice that, when using the application, this configuration actually appears at a different position, namely under the 'Properties' header in ONDA. The screenshot has been modified for presentation simplicity in this paper. After clicking on the 'Enrol1' line, the relationship can be configured in the properties pane. We can configure the cardinality of the relationship at both ends of the relationship, with the different symbols following the Crow's Foot notation. In the case of the example shown in Figure 2, a Student may perform none or various enrollments, one enrollment belongs exactly to a single student. In this particular case, Enrollment is also a \textbf{Weak Entity}, which means it either does not have an identifier or it has a partial identifier only, which is the case of the example. We can tick the 'is weak' checkbox to identify this type of situation.

Finally, two entities may have an \textbf{hierarchical relationship}, which is the case of Person (the super-entity) and Professor and Student (the sub-entities). This type of relationship can be configured to 'Complete', 'Concrete', or 'Single' which essentially result on the generation of different physical diagrams. The 'Complete' configuration generally means that all sub-entities will have direct correspondence to a table in the physical diagram. 'Concrete' will result in distribution of the attributes of the super-entity by the sub-entities and in the removal of the table corresponding to the super-entity. Finally, the 'Single' configuration results in merging the attributes of the super-entity and of the sub-entities in a single table in the physical model.

\subsection{Physical Model Generation}

Figure 3 shows the physical data model that corresponds to the ER diagram shown earlier.

\begin{figure*}[!h]
    \centering    \includegraphics[width=0.8\linewidth]{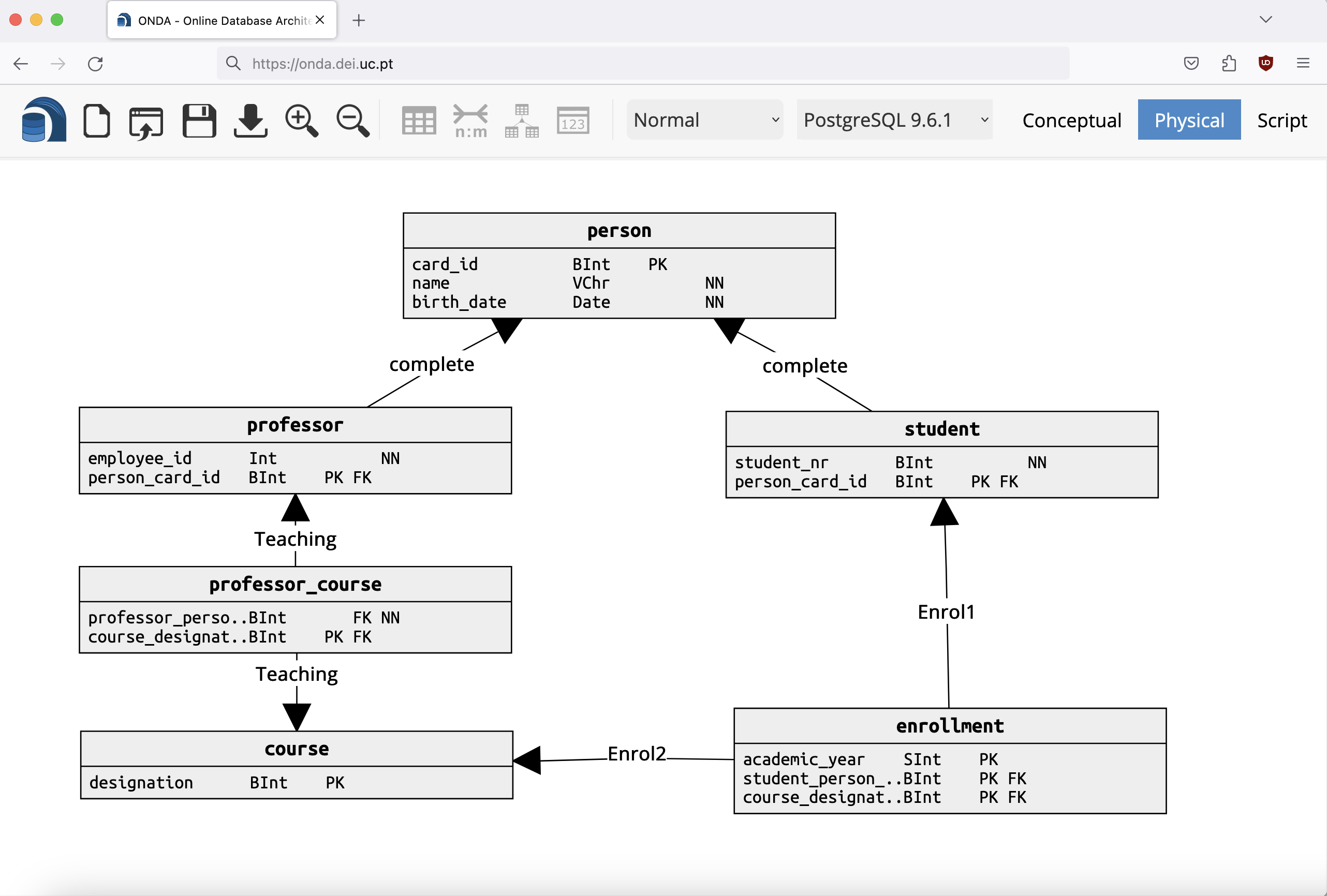}
    \caption{Physical data model generation in ONDA.}
\end{figure*}

There are basically two alternatives for configuring the physical data model generation: 'Normal' and 'Simplified', which can be selected from the drop-down list visible at the top of Figure 3. In this case, the option selected is 'Normal', which has the main effect of generating one additional table \texttt{professor\_course} where the associations between professors and courses are to be stored. This intermediate table does not admit null values and this is the main idea behind the 'Normal' generation mode. In the generated table \textit{professor\_course}, we can see that a given course can only appear once (\texttt{course\_designation} is primary key) and a given employee (referenced by its card identifier) can appear multiple times. This means that the generated table will allow for an employee to be associated with different courses, however a given course cannot be associated with more than one employee, which is what has been specified in the previous conceptual model shown in Figure 2.

\subsection{SQL Generation}

Figure 4 shows the SQL generated by ONDA, corresponding to the physical data model presented in the previous subsection.

\begin{figure*}[!h]
    \centering    \includegraphics[width=0.8\linewidth]{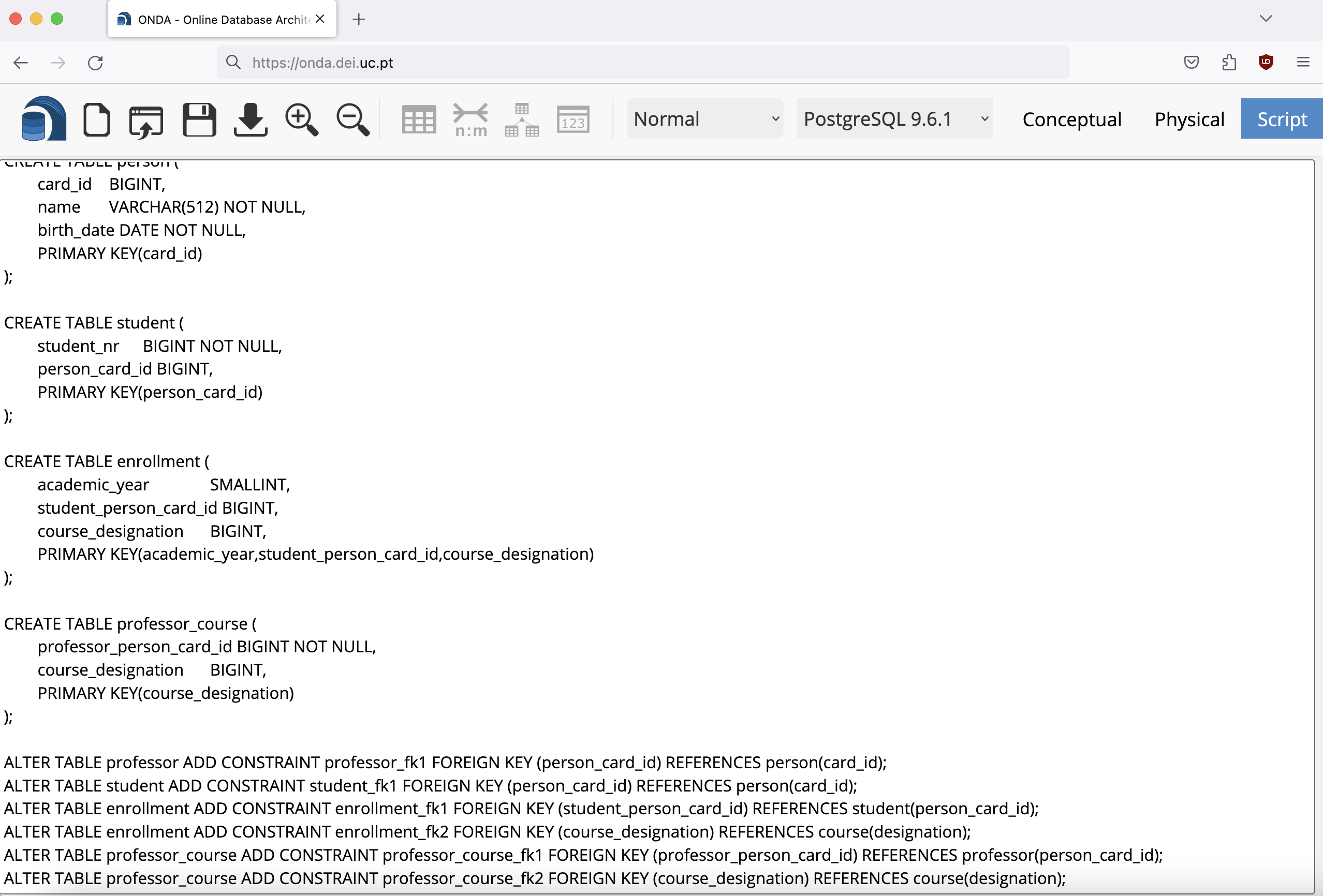}
    \caption{SQL generation in ONDA.}
\end{figure*}

In the case of SQL generation, the only configuration possible is the selection of one out of five different database engines, namely PostgreSQL, Oracle, MySQL, MariaDB, SQLite. It is worthwhile mentioning that PostgreSQL is the engine for which SQL generation has been most thoroughly tested.

The generated SQL includes instructions for the creation of the tables along with the necessary constraint definitions (e.g., foreign keys, mandatory columns). We aimed at creating standard SQL, although in certain circumstances the different database management systems require some customization of the generated SQL.
\section{Conclusion}

This paper presented ONline Database Architect (ONDA), a web-based application that allows users to create Entity-Relationship diagrams, generate the corresponding physical data models, and the corresponding SQL. ONDA is, at the time of writing, being used by more than three hundred students every academic year. Our future plans include extending the tool to support further modeling notations or allowing it to edit and store PL/SQL code.

\begin{acks}

The main driver behind the development of ONDA was the lack of free, cross-platform, easy to use, applications that for drawing Entity-Relationship diagrams. We aimed at a tool that could generate the physical diagram and also the corresponding SQL, which would be particularly tailored for a university-level learning environment. 

We would like to thank the various students involved in the development of the application throughout the years. We would also like to thank the 'Project Management' course offered in the Master in Informatics Engineering at the University of Coimbra, which allowed three teams of students to participate in this project. Also we thank our colleague Marco Vieira for various key suggestions in the initial stage of development of ONDA, many years ago. The current version of ONDA (v4) has been polished with the help of one additional student, for which we involved our colleague João Campos who provided guidance for the work and to whom we extend our thanks.

This work has been supported by the FCT – Foundation for Science and Technology, I.P./MCTES through national funds (PIDDAC), within the scope of CISUC R\&D Unit – UIDB/00326/2020 or project code UIDP/00326/2020

\end{acks}

\vspace{-0.21cm}

\bibliographystyle{ACM-Reference-Format}
%\bibliography{sample-base}
\bibliography{library}

\end{document}